\documentclass[twocolumn,secnumarabic,amssymb, nobibnotes, aps, prc, superscriptaddress, nobalancelastpage,longbibliography,nofootinbib]{revtex4-1}

\usepackage{graphicx}
\usepackage{amsfonts}
\usepackage{amsmath,amssymb}
\usepackage{bm}
\usepackage{xcolor}
\usepackage{float}
\usepackage{dcolumn}
\usepackage{cancel}
\usepackage{ulem}

\usepackage{subfig}
\usepackage{booktabs}
\usepackage{physics}
\usepackage{multirow}
\usepackage{bm}
\usepackage{appendix}
\usepackage{xr-hyper}
\usepackage{hyperref}
\hypersetup{breaklinks=true,colorlinks=true,linkcolor=blue,citecolor=blue,filecolor=magenta,urlcolor=cyan}

\begin{document}

\title{Microscopic calculation of the pinning energy of a vortex in the inner crust of a neutron star}

\author{P. Klausner}
\affiliation{Dipartimento di Fisica ``Aldo Pontremoli'', Universit\`a degli Studi di Milano, 20133 Milano, Italy}
\email{pietro.klausner@unimi.it}

\author{F. Barranco}
\affiliation{Departamento de F\'isica Aplicada III, Escuela Superior de Ingenieros, Universidad de Sevilla, Camino de Los Descubrimientos, Sevilla, Spain}
\email{barranco@us.es}

\author{P. M. Pizzochero}
\affiliation{Dipartimento di Fisica ``Aldo Pontremoli'', Universit\`a degli Studi di Milano, 20133 Milano, Italy}
\affiliation{INFN, Sezione di Milano, 20133 Milano, Italy}
\email{pierre.pizzochero@mi.infn.it}

\author{X. Roca-Maza}
\affiliation{Dipartimento di Fisica ``Aldo Pontremoli'', Universit\`a degli Studi di Milano, 20133 Milano, Italy}
\affiliation{INFN, Sezione di Milano, 20133 Milano, Italy}
\email{xavier.roca.maza@mi.infn.it}

\author{E. Vigezzi}
\affiliation{INFN, Sezione di Milano, 20133 Milano, Italy}
\email{vigezzi@mi.infn.it}

\date{\today}
\begin{abstract}
The structure of a vortex in the inner crust of a pulsar is calculated microscopically in the Wigner-Seitz cell approximation, simulating the conditions of the inner crust of a cold, non-accreting neutron star, in which a lattice of nuclei coexists with a sea of superfluid neutrons.
 The calculation is based on the axially deformed Hartree-Fock-Bogolyubov framework, using effective interactions. The present work extends and improves previous studies in four ways: 
 i) it allows for the axial deformation of protons induced by the large deformation of neutrons due to the appearance of vortices; 
 ii) it includes the effect of Coulomb exchange; 
 iii) considers the possible  effects of the screening of the pairing interaction;
 and iv) it improves the  numerical treatment. 
 We also demonstrate that the binding energy of the nucleus-vortex system can be used as a proxy to the pinning energy of a vortex and discuss in which conditions this applies. 
 From our results, we can estimate the mesoscopic pinning forces per unit length acting on vortices. 
 We obtain values ranging between  $10^{14}$ to $10^{16}$ dyn/cm, consistent with previous findings.
\end{abstract}

\maketitle



\section{Introduction} \label{sec:introduction}

Pulsars are characterized by the regular emission of electromagnetic radiation.
These stars  spin down steadily, but sudden spin-ups, called glitches, have been observed.
Such events were  recorded first in the Vela pulsar and subsequently in many other stars (see \cite{fuentes17} for a statistical study of the properties of glitches observed in 141 stars). 
Soon after the first observations, it was proposed that the glitch phenomenon was closely associated with the existence of a neutron superfluid in the interior of the star \cite{baym1969Nat}, see \cite{Haskell2015IJMPD,AMP_review_2023} for a review. 
According to the current theoretical understanding of neutron star structure, the layer extending from a density of about $10^{-3}$ fm$^{-3}$ to 0.04 fm$^{-3}$, called the inner crust, is composed of a lattice of heavy nuclei immersed in a sea of free neutrons and electrons \cite{chamel2008LRR,NS1book}.
Negele and Vautherin carried out a seminal study \cite{negele1973} within  the Wigner-Seitz approximation.  They determined the optimal radius of a spherical cell with a nucleus at its center, the number of protons and of neutrons  bound to the nucleus and the number of unbound neutrons, as a function of the neutron density at the edges of the cell.
Their results have been refined and extended in many subsequent works, see \cite{pizzochero2002,sandulescu2004,baldo2007,grill2011,pastore2011,mondal,shelley2021} and references therein. Moreover, given the typical range of temperature expected in the inner crust of mature neutron stars (from $10^7$ to $10^9$ K, that is from 1 to 100 keV, a very low value with respect to the Fermi energy ranging from 10 to 100 MeV), neutrons are likely to be superfluid \cite{sedrakian2019EPJA}.

Due to the rotation of the star, the superfluid neutrons form a (possibly disordered) array of quantum vortices \cite{Feynman}, whose average density is closely linked to the pulsar angular velocity via a generalization of the so-called Feynman-Onsager relation \cite{AMP_mnras_2018}.
Anderson and Itoh  \cite{Anderson&Itoh} proposed that  the interaction between the heavy nuclei at the lattice sites and the vortices can anchor the vortices in particularly energetically favourable positions, a phenomenon referred to as ``pinning''. 
If this is the case, the superfluid component cannot  follow the regular slowdown of the crust and rotates faster, becoming a reservoir of angular momentum. 
This gives rise to hydrodynamical lift forces (Magnus forces), which act on the vortex lines and tend to push them away from their sites. 
The glitch phenomenon would then occur when Magnus forces take over and a catastrophically large number of vortices suddenly unpin from their positions, releasing  their  angular momentum to the crust.

There are still some unanswered questions regarding several central aspects of this model.
First of all, the trigger which leads to the collective vortex unpinning  is not well established yet; there are several possibilities advanced in the literature, like vortex avalanches \cite{Anderson&Itoh,vortexAvalnches} or hydrodynamical instabilities \cite{hydInstabilities,Instabilities2019PhRvD}.
Secondly, it has been pointed out that the angular momentum contained in the crust may not be sufficient \cite{crustNotEnough,crustalEntrainment} to explain glitches, albeit this conclusion is less clear if the statistical uncertainty on the observed glitch activity \cite{MAP2021Univ} or the possible presence of lattice defects \cite{WonderfulPaperBySauls_arXiv_2020} are taken into account.
Finally, there is no definitive answer on the strength of the pinning interaction throughout the inner crust.
The greater  the ability of pinning  to  withstand the hydrodynamical lift, the higher the amount of  angular momentum that the superfluid can store, so that it is possible to constrain the unpinning threshold (i.e. the theoretical upper limit of the distribution of pinning forces \cite{AH_2020MNRAS}) with observations of large glitches~\cite{AMP_mnras_2018}.

The microscopic computation of the single-nucleus pinning potential is very challenging and  has never been performed in the literature.
In fact, existing studies resorted to the pinning energy \cite{pierre2, Avogadro&Al, epstein&baym1988}, defined as the energy difference between two extreme situations: one where the vortex is on top of the nucleus (nuclear pinning), and one where the vortex is equidistant between two adjacent nuclei in the lattice (interstitial pinning).
A negative (positive) value of this quantity indicates that the former (latter) situation is energetically favourable.

Different  methods have been used to estimate the single-nucleus pinning potential.
Epstein and Baym ~\cite{epstein&baym1988} used hydrodynamic considerations in combination with the Ginzburg–Landau theory of superfluidity to compute the free energy of a nucleus as a function of the distance from a vortex line, ignoring the internal structure of the nucleus and using instead schematic expressions for the kinetic and condensation energies. 
They found that  vortices pin on nuclei in the deeper layers of the inner crust, while they are repelled in the low-density regions. 
 The model by Epstein and Baym was later improved \cite{pierre1}, providing estimates of pinning energies obtained by making use of a semiclassical treatment  based  on the Local Density Approximation \cite{pierre2}.
 
The first  microscopical quantum calculation 
was then carried out by \citet{Avogadro&Al,Avogadro&Al2},
based on the solution of the axially symmetric  Hartree-Fock-Bogoliubov (HFB) equations in the Wigner-Seitz approximation for various densities in the crust.
Specifically, it was found that the nuclear shell structure has relevant effects on the spatial configuration of the vortex and that pinning occurs only in the less dense regions of the inner crust. 
The solution of the HFB equations was carried out assuming spherical symmetry for the proton density, thus breaking self-consistency.  
In the present paper, we remove this assumption, which was based on the fact that proton orbitals are deeply bound.  
Furthermore, we include the effect of the Coulomb exchange, which was previously neglected, and improve the numerical treatment, devoting particular attention to  the convergence of our results. 
We are then able to present new and more reliable values of the binding energy and, based on  them, we present our best estimation of the pinning energy.
We also show detailed results for neutron and proton deformation at different densities.
We also study the dependence of our results on the strength of the pairing interaction, in keeping with the analysis carried out in \cite{pierre2}. 

Due to the fact that hydrodynamics is non-linear, the pinning potential is not immediately related to the pinning ``landscape'' that defines the dynamics of a finite-size vortex segment \cite{AH_2020MNRAS}.
We then estimate  the typical strength of the pinning landscape by taking the  mean value of the pinning force for unit length acting on a vortex line \cite{PMP2016}, see also the discussion in~\cite{AH_2020MNRAS}.

Other recent efforts, based on a microscopic quantal picture, have also been made. 
The most significant advance concerns a  three-dimensional dynamical simulation of the vortex motion, based on the  time-dependent superfluid local density approximation (TDSLDA), leading to an estimate of the force between the vortex and the nucleus as a function of their separation \cite{walzlowski2016,bulgac2013}(see also \cite{link2009}). 
Results were obtained for two densities and showed that the vortex is repelled by nuclei. 
At the same time, it was found that the vortex-nucleus interactions induce a deformation of the nucleus and lead to a bending of the vortex line shape.
These findings  represent an important confirmation of our results and extend them toward a complete characterization of the vortex-nucleus interaction. 
On the other hand, TDSLDA computations are very costly, while we are able to present systematic calculations of the pinning energy with different functionals and pairing forces and to provide a detailed description of the nuclear deformation. 
We also report that the properties of a quantum vortex were recently studied at finite temperature in infinite matter using Brussels-Montreal energy functionals \cite{pecak2021}. 

We begin in Section \ref{sec:method} by explaining the general features of the calculation and giving some details about the computation of the pinning energy. Our results are presented in Section \ref{sec:results}.
Finally, in Section \ref{sec:conslusions} we give our closing remarks.

\section{Method} \label{sec:method}
\subsection{General description}
In this paper, we expand and improve the work done in \cite{Avogadro&Al} (hereafter referred to as Paper I).
There, the authors approached the problem of pinning energy by solving the Hartree-Fock-Bogolyubov (HFB) equations in a cylindrical Wigner-Seitz cell of radius $R_{WS}$ and height $h_{WS}$ in four different configurations. 
HFB equations (also called Bogliubov-De Gennes equations) are well suited to study  the pairing properties of quantal inhomogeneous systems, like the inner crust of a neutron star, where a lattice of heavy nuclei coexists with a sea of superfluid neutrons.
With this technique both the nuclear potential and the pairing correlations are treated simultaneously and self consistently.
Explicitly, the HFB equations read
\begin{equation}
        \begin{cases}
        \left(h(\mathbf{x}) - \lambda\right)u_{i}(\mathbf{x}) + \Delta(\mathbf{x})v_{i} (\mathbf{x}) = E_{i} u_{i}(\mathbf{x}) \\
        \Delta^*(\mathbf{x}) u_{i}(\mathbf{x}) - \left(h(\mathbf{x}) - \lambda\right)v_{i} (\mathbf{x}) = E_{i} v_{i}(\mathbf{x})
    \end{cases}
    \label{eq:hfbCoordinate}
\end{equation}
where $E_i$ is the quasi-particle energy of level $i$ and $u_{i}$ and $v_{i}$ are the quasi-particle amplitudes relative to that level, $\lambda$ is the chemical potential, $\Delta(\mathbf{x})$ is the pairing field and $h(\mathbf{x}) = T + U^{HF}$ is the single particle Hartree-Fock Hamiltonian, sum of the kinetic term $T$ and the self-consistent potential $U^{HF}$.

From the solutions of (\ref{eq:hfbCoordinate}), one can compute the normal and abnormal densities of the system
\begin{equation}
    \begin{split}
        n(\mathbf{x}) = \displaystyle\sum_{i}|v_{i}(\mathbf{x})|^2 \\
        \kappa(\mathbf{x}) = \displaystyle\sum_{i}u_{i}(\mathbf{x}) v_{i}(\mathbf{x})^*
    \end{split}
    \label{eq:density}
\end{equation}
from which one can find new $h(\mathbf{x})$ and $\Delta(\mathbf{x})$ which in turn give rise to a new set of equations (\ref{eq:hfbCoordinate}) (see Appendix A). 
The HFB equations are therefore solved via an iterative process.

As for the interaction chosen in the HF sector, we adopt the Skyrme SLy4 and the SkM* parameterizations
 (see \cite{chabanat1}) and neglect the spin-orbit term, because we expect that  the pinning energy is not significantly affected by this term (cf.  Paper I and our discussion below).

For the pairing field, we start from a neutron pairing potential, adopting a density-dependent, contact interaction of the  form
\begin{equation}
    V_{pair}(\mathbf{x},\mathbf{x}') = V_0 \left(1 -\eta \left(\frac{n(\mathbf{x})}{0.08}\right)^a\right) \delta(\mathbf{x}-\mathbf{x}')
    \label{eq:pairPotential}
\end{equation}
where $V_0 = -481$ MeV$\,\cdot \,$fm$^{3}$, $\eta=0.7$ and $a=0.45$ have been used.
This leads in turn to the pairing field
\begin{equation}
    \Delta(\mathbf{x}) = - V_{pair}(\mathbf{x},\mathbf{x}') \kappa(\mathbf{x})
    \label{eq:pairingField}
\end{equation}
The adopted parameters, together with a cutoff energy $E_{cut}= $ 60 MeV, reproduce the pairing gap of uniform neutron matter as predicted by a realistic nucleon–nucleon interaction \cite{pairingComparison}, and are the same as those used in Paper I.
We will also perform calculations with two weaker pairing interactions. 
We aimed for pairing gaps reduced by a factor $\beta=2$  and $\beta=3$; we found  $V_0^{\beta=2} = 432.9$ MeV$\,\cdot \,$fm$^{3}$ and $V_0^{\beta=3} = 408.85$ MeV$\,\cdot \,$fm$^{3}$.
These interactions are introduced only to have a rough qualitative assessment of the effects of correlations beyond the mean field, which generally lead to a reduction of the pairing gaps (see \cite{Ramanan2021,Gandolfi2022} for recent reviews).
However, such reductions show a dependence on the neutron density which is not taken into account  by the constant reduction factors considered here.
Nonetheless, we will still label the results by $\beta=2$ and $\beta=3$.

The pairing interaction has been neglected in the case of protons since $Z$=40 is used throughout this work and this value corresponds to a magic number in our calculations.  

We carry out our calculations in a cylindrical box, so it is natural to use cylindrical coordinates $\mathbf{x} = (\rho,z,\varphi)$.
Eqs. (\ref{eq:hfbCoordinate}) are  expanded  on a single-particle basis. 
All the calculation details are presented in Appendix \ref{app:ND}.
The pairing field (\ref{eq:pairingField}) is defined as (Paper I and \cite{pairingVortex})
\begin{equation}
    \Delta(\rho,z,\varphi) = \Delta(\rho,z)\,e^{i\nu\varphi}
    \label{eq:pairingVortex}
\end{equation}
so that the vortex is created along the $z$-axis keeping the cylindrical symmetry. 
The integer parameter $\nu$ can be interpreted as the number of  units of angular momentum  carried by each Cooper pair along the $z-$axis.  
The standard solution of the HFB equations corresponds to $\nu = 0$ and to  Cooper pairs coupled to zero angular momentum   while $\nu = 1$ defines an excited solution in which Cooper pairs of different parity couple to one unit of angular momentum.
This solution describes a vortex, as it gives rise to an azimuthal velocity field $V$ of the form
\begin{equation}
    V(\rho,z,\varphi) = - \frac{i\hbar}{m n \rho} \displaystyle\sum_i v_i^*(\rho,z,\varphi) \frac{\partial v_i(\rho,z,\varphi)}{\partial \varphi}.
    \label{eq:velocityField}
\end{equation}
It is noted that  nuclear shell effects act quite differently on the $\nu=1$ gap, as compared to $\nu=0$.  
This  point is discussed at length in Paper I. 
In particular, one expects that the spin-orbit interaction, which is neglected in the present work, tends to shift the energy of the single-particle pairs involved in the formation of $S=0, \nu=1$ Cooper pairs by the same amount (see Fig. 21 in Paper I).

We have changed considerably the part of the computation relative to protons with respect to Paper I. 
In Paper I, the proton density was forced to be spherically symmetric. 
This was achieved by taking spherical averages of the cylindrical neutron densities to compute the proton potential $U^{HF}_{prot}$ at each step of the iterative process. 
The reasoning behind this choice was that protons are deeply bound and one does not expect them to be much affected by the neutron density deviation from sphericity. 
As we will show, this is an accurate approximation only for the outermost layers of the inner crust.



Summarizing, we have extended and improved the calculations of Paper I as follows:
\begin{itemize}
    \item we add the Coulomb exchange term in the proton potential using the Slater approximation.  
    \item we adopt cylindrical symmetry also in the case of  protons.
    \item we consider, although schematically, the effects associated with the possible reduction of the pairing interaction due to screening effects.
    \item we improve the numerical aspects of the code, namely the derivation and integration techniques.  Improving the numerical precision is crucial for computing the pinning energy, as we will show in the next section.

\end{itemize}


\subsection{Binding and pinning energy} \label{sec:method.pinningEnergy}
\begin{figure*}
    \centering
    \includegraphics[width=1\textwidth]{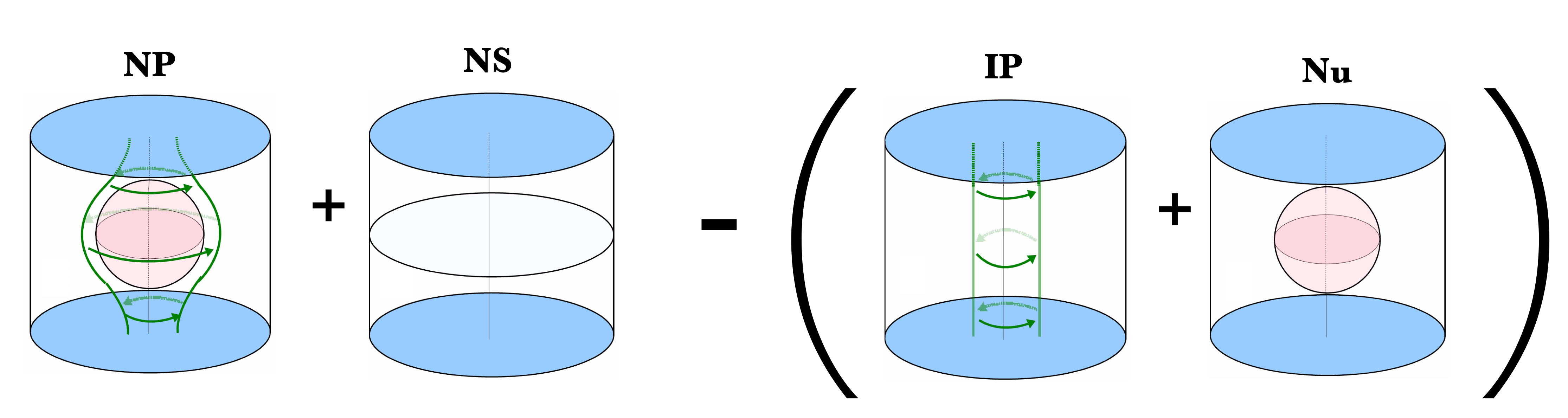}
    
    \caption{Visual representation of \eqref{eq:bindingEnergyTheoric}. The binding energy is shown as the energy cost to move a vortex from its position on top of a nucleus to an infinite distance from it.}
    \label{fig:bindingEnergy}
\end{figure*}

We solve the HFB equations in the following configurations (see  Fig.~\ref{fig:bindingEnergy} for a sketch):
\begin{itemize}
    \item {\bf Neutron sea (NS):} the neutron sea, with neither a nucleus ($Z=0$) nor a vortex ($\nu=0$); 
    \item {\bf Nucleus (Nu):} a nucleus ($Z\neq 0$) with no vortex ($\nu=0$), surrounded by the neutron sea;  
    \item {\bf Interstitial pinning (IP):}  a vortex ($\nu=1$) with no nucleus ($Z=0$), surrounded by the neutron sea;
    \item {\bf Nuclear pinning (NP):}  a nucleus ($Z\neq 0$) and a vortex ($\nu=1$) on top of it, surrounded by the neutron sea.
\end{itemize}

By comparing the total energies of each configuration, we computed the \textit{binding} energy of the vortex onto the nucleus.
This quantity is defined as the difference between the energy needed to build a vortex on top of a nucleus and the energy necessary to build a vortex in uniform matter.
Equivalently, the binding energy can be defined as the energy needed to move the vortex from its site on top of the nucleus to an infinite distance from it (see Fig. \ref{fig:bindingEnergy}). 
A negative value means that the favorable position for the vortex is on top of the nucleus, whilst a positive value means that the favorable position is far away from it.

A simple combination of the total energies of each configuration gives the explicit expression of the binding energy
\begin{eqnarray}
    E_b &=& E^{NP} + E^{NS} - (E^{IP} + E^{Nu}) \nonumber\\ &-& \lambda_n \left[ (N^{NP} + N^{NS} -(N^{IP} + N^{Nu})\right]
    \label{eq:bindingEnergyTheoric}
\end{eqnarray}
where $E^i$ is the total energy of the specified configuration. 
We added a correction term proportional to the neutron chemical potential $\lambda_n$ to ensure that we compare calculations with the same number of particles, since the vortex, if present, reduces the number of neutrons $N^i$ found in each cell.

Numerical precision is crucial to compute the binding energy.
The energy terms in \eqref{eq:bindingEnergyTheoric} range from some hundreds of MeVs up to  tens of thousands MeVs
as a function of neutron density in the inner crust.
The values of the nucleus-vortex binding energy, on the other hand, range from some hundreds of keVs up to tens of MeVs.
Even small numerical errors can have substantial effects on the final values of the binding energy.

The binding energy is a different quantity with respect to the \textit{pinning} energy $E_p$.
The latter is influenced by the presence of the surrounding nuclear lattice and therefore we are unable to calculate it directly.
Nonetheless, we can find an estimate through the binding energy.

Epstein and Baym in \cite{epstein&baym1988} realized that there is a kinetic component to the vortex-nucleus interaction, that accounts for the amount of superfluid flow displaced by the nucleus.
It reads  
\begin{equation}
    K_n(\rho) = \frac{3}{2} M_s \left(\frac{\zeta -1}{\zeta + 2}\right) \left(\frac{\hbar}{2m_0\rho}\right)^2
    \label{eq:kineticTermPE}
\end{equation} 
where $m_0$ is the nucleon mass, $M_s$ is the mass of the neutron superfluid of density $n_\infty$ displaced by a sphere of radius $R_n$ (i.e., the nuclear radius) and $\zeta$ is the ratio of the nucleus density $n_n$ to the neutron superfluid density $n_\infty$.
$K_n$ is always positive and it is inversely proportional to the square of the distance $\rho$ between the nucleus center and the vortex axis.

On the other hand, the other component of the interaction is of nuclear nature.
If we assume that such nuclear interaction is short-ranged, then after a certain critical distance $\rho^*$ it will become negligible, along with its contribution to the pinning energy.
We can estimate such distance as the sum of the nuclear radius $R_n$ and the coherence length $\xi$ of the vortex
\begin{equation}
    \rho^* \sim R_n + \xi
    \label{eq:rhoCrit}
\end{equation} 
where $ \xi = \hbar^2 k_F/\pi m_0 \Delta$, with $k_F$ the Fermi momentum. 
From our calculations, $\xi$ ranges between 3 and 10 fm approximately, depending on the density of the neutron sea.

To compute the pinning energy, we must compare $\rho^*$ with $R_{WS}$.
We assume that the nuclear contribution to the vortex-nucleus interaction is negligible for $\rho  \gtrsim \rho^*$.
If $\rho^*<R_{WS}$, we then suppose that at $\rho = R_{WS}$ the vortex-nucleus interaction is dominated by the kinetic term \eqref{eq:kineticTermPE}.
Therefore, from the definition of pinning energy, we write
\begin{equation}
    E_p \simeq E_b - K_n(R_{WS})
    \label{eq:pinningEnergyExplicit}
\end{equation} 
At $R_{WS}$, the contribution of $K_n(R_{WS})$ is of the order of a few tens of keV, so that it usually  
represents  a small correction to the pinning energy.

If, on the contrary, $\rho^*\gtrsim R_{WS}$,  there would still be 
a substantial overlap between the vortex and the  nucleus at a  distance $\rho = R_{WS}$. 
In this case, we are unable to estimate the non-negligible nuclear component to the interaction and therefore we cannot provide an estimate on the pinning energy.

\subsection{Computational details} \label{ssec:computations_details}
Similarly to Paper I, we present the calculated value  of the pinning energy as a function of the density of the  neutron sea far from the nucleus, $n_\infty$.
We investigated eight different density zones, from $n_\infty=0.001$ fm$^{-3}$ to $n_\infty=0.038$ fm$^{-3}$.
At each density, we have carried out six sets of calculations, using two different Skyrme models, namely SLy4 and SkM*, and three different pairing strengths (marked by the pairing-interaction reduction factor $\beta$). For each set, we iteratively solved two  HFB equations, one for protons and one for neutrons, for each of  the four different configurations.

The neutron chemical potential was chosen so as  to reproduce  the external densities predicted in \cite{negele1973} and studied in Paper I.
On the other hand, the proton chemical potential was adjusted to give the proton number $Z=40$ \cite{negele1973}. 

We took special care in estimating the errors due to the convergence of the calculations and also those due to the size of the box, which is essential for our results to be reliable. 
Specifically, we adopted the following  convergence criterion for the computation of a given configuration: the program halts when the relative total energy difference between the last and second-last iteration is less than $5\times 10^{-6}$ for three consecutive iteration cycles. 
In some cases, we observed that this criterion was not stringent enough; we let therefore the computation continue until the relative energy difference reached $5\times 10^{-8}$ for three consecutive iteration cycles.


After the binding energy was obtained, we  computed the critical distance $\rho^*$ \eqref{eq:rhoCrit} as well as the kinetic contribution \eqref{eq:kineticTermPE} (which within our approximation does not depend on the box radius).
If the criterion $\rho^* < R_{WS}$ was met, we were able to compute the corresponding pinning energies via \eqref{eq:pinningEnergyExplicit}; otherwise, we concluded that our method could not produce a result for the particular parameter set.
In Appendix \ref{app:rhoStar} we show the values of $\rho^*$ we obtained.



\section{Results}
\label{sec:results}

\subsection{Vortex effects on pairing gaps and proton deformation}

\begin{figure*}
    \includegraphics[width=1\linewidth]{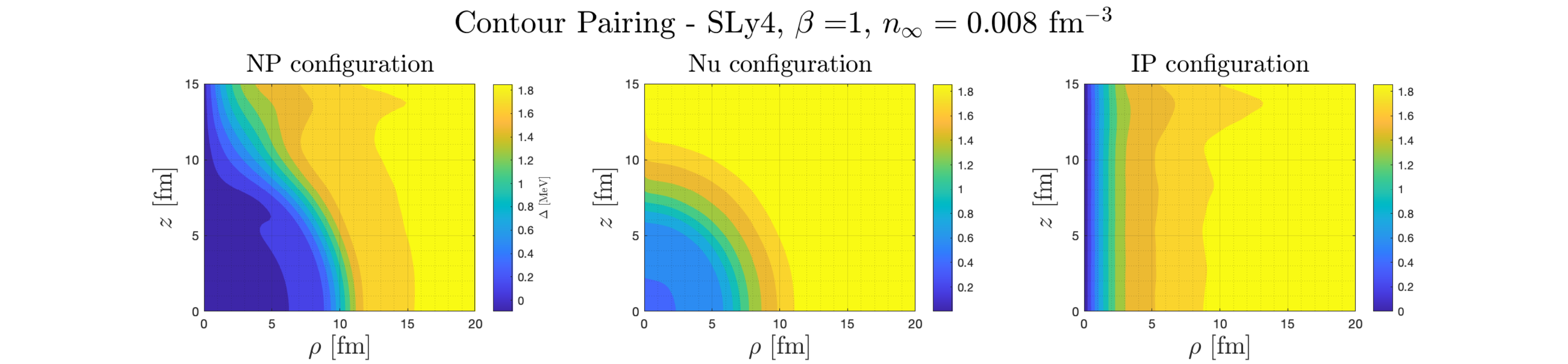}
    \caption{Contour plots of the pairing gaps of the NP (left), Nu (center), and IP (right) configurations.}
    \label{fig:contour_pairing}
\end{figure*}

\begin{figure*}
    \includegraphics[width=.75\linewidth]{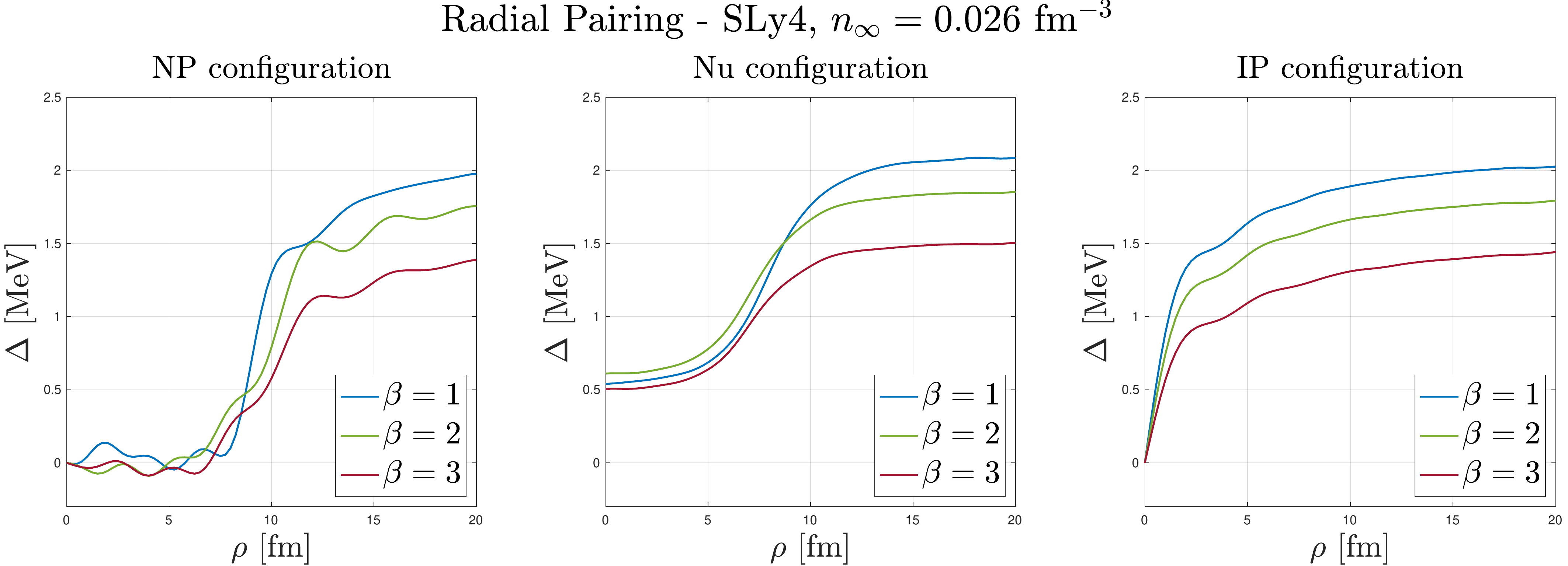}
    \caption{Typical pairing gaps obtained in our calculations for the NP,  Nu, and IP configurations, for the SLy4 interactions, and for the three adopted values of $\beta$, as a function of the distance from the vortex axis in the $z=0$ plane.}
    \label{fig:radial_pairing}
\end{figure*}
In Fig.~\ref{fig:contour_pairing} we compare contour plots of the pairing gaps associated with the NP  (left), Nu (center), and IP (right) configurations in the $(\rho,z)$ plane, calculated with the SLy4 interaction for the density $n_{\infty} = 0.008$ fm$^{-3}$. One can see that the gap  acquires its asymptotic value for $\rho \gtrsim $ 10 fm in the IP configuration, while the presence of the nucleus distorts the gap profile in the NP configuration so that the vortex enlarges and incorporates the nucleus, and the gap reaches its  asymptotic value only for $\rho \gtrsim $ 15 fm. 
Our results are qualitatively consistent with those obtained in \cite{walzlowski2016}, where the vortex-nucleus interaction was studied with dynamical simulations (see  Fig. 2 in \cite{walzlowski2016}, where one can actually observe the vortex bending to avoid the nuclear region). 
The gap profiles for the NP, Nu, and IP configuration along the equator $z=0$ are shown in Fig. \ref{fig:radial_pairing} for the SLy4 interaction and the three values of $\beta$ we have considered.
The density is $n_{\infty} = 0.026$ fm$^{-3}$. In all cases, the gap is suppressed for $\rho \leq $ 10 fm and rapidly reaches the asymptotic value corresponding to the given value of $\beta$. There is a slight dependence on the interaction, which essentially depends on the different values of the effective mass associated with  the SLy4 and with  the SkM$^*$ interaction. 

\begin{figure*}
    \hspace*{-2cm}
    \includegraphics[width=1.2\linewidth]{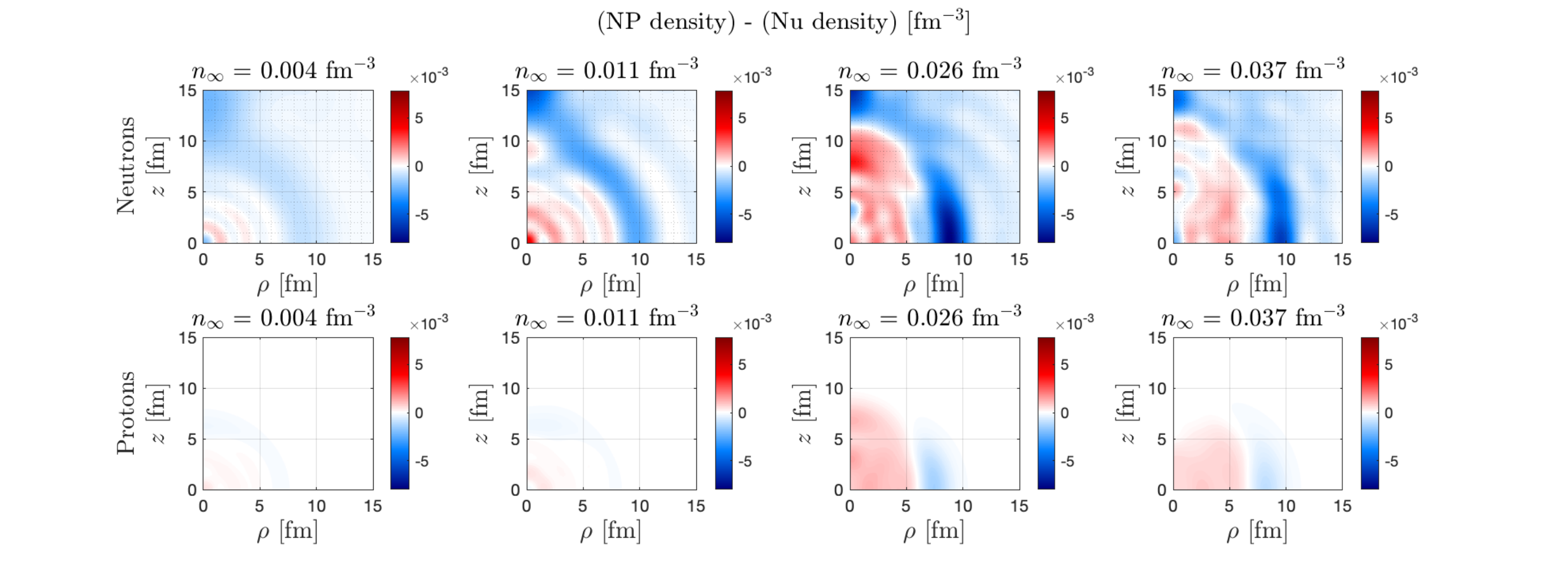}
    \caption{ Difference between the densities calculated in  the NP and Nu configurations, expressed in fm$^{-3}$, as a function of ($\rho, z$) in a $\varphi$-constant plane for several neutron sea densities $n_\infty$. 
    In the top half, we show neutron quantities, while in the bottom half proton quantities.}
    \label{fig:deformation}
\end{figure*}


In Fig. \ref{fig:deformation} we present contour plots in the ($\rho,z)$ plane of the differences between the  density distributions calculated in the NP and in the Nu  configuration  with the SLy4 interaction (see also \cite{Avogadro&Al3}).
Upper and lower panels  refer to neutrons and to protons respectively. 
We have set the same color scale for both neutrons and protons and  we display results obtained for four different Wigner-Seitz cells corresponding to varying depths in the inner crust.
Deformation effects increase as a function of density. 
The deformation of the nucleus tends to be prolate, that is, aligning the nuclear density with the axis of the vortex.
In the neutron case,  it is possible to observe a density depletion (circular blue shadow) surrounding the nucleus ($\rho \lesssim 7$ fm and $z \lesssim 7$ fm).
This is an expected  effect of the internal regions of a fermionic vortex (see Paper I  and \cite{pecak2021} for more details), that takes place at all densities and for the three  $\beta$ factors. 
The only exceptions are found in the case of the  SkM* interaction where one observes some penetration of the vortex into the nucleus at the two highest neutron sea densities  (not shown in the figures).

In general, the deformation of the distribution of protons is similar in shape and magnitude to that of neutrons (giving rise to variations in the density up to 5-10\% in the case of high-density cells). 
This can be considered to be the result of the general tendency of the nucleus to maximize the overlap between the distribution of neutrons and protons. We will assess the effect of the deformation on pinning energies below.


It is reasonable to think that  this trend should continue as we move to deeper and denser areas of the crust, where the pasta phase will most likely produce negative pinning energy, thus giving rise to a hitherto unexplored hybrid mode of pinning.

Hence, the vortex-nucleus interaction may favor the appearance of the pasta phase, thought to be present at higher densities than the ones studied here  \cite{pearson2022}. 
Moreover, the appearance of the nuclear pasta is expected to influence the pinning interaction, with consequences for the macroscopic hydrodynamic behavior of the superfluid in the pasta layers \cite{AH_2020MNRAS}. 
This interesting subject is left for future studies. 
The effect of deformation on the pinning energy will be discussed in the next section.

\subsection{Pinning Energies}

\begin{figure*}
    \hspace*{-2cm}
    \includegraphics[width=1.2\textwidth] {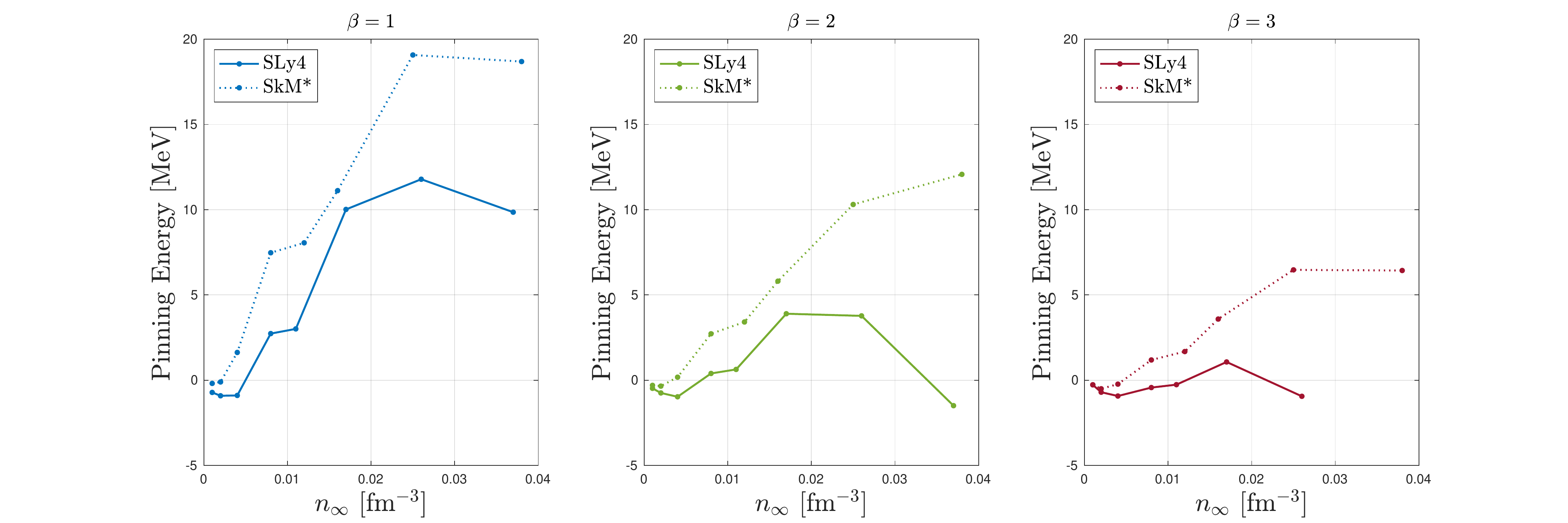}
    \caption{Pinning energies as a function of the neutron sea density $n_\infty$, for three values of $\beta$ and for both SLy4 (straight line) and SkM* (dotted line) interactions.
    The highest density point with SLy4 and $\beta=3$ is absent because it does not satisfy our requirement $\rho^* > R_{WS}$ (see section \ref{sec:method.pinningEnergy}).}   
    \label{fig:pinningEnergy}
\end{figure*}
In Fig. \ref{fig:pinningEnergy} we show our results for the pinning energy as a function of  the neutron sea density $n_\infty$ for both SLy4 (straight line) and SkM* (dotted line) interactions. The corresponding numerical values are reported in Tab. \ref{tab:pinningEnergiesSly} and \ref{tab:pinningEnergiesSkm}. 

The value of the pinning energy depends considerably on the value of the interstitial pairing gap, which could be much lower than the bare gap (especially at high densities) due to screening effects. 
For this reason, we have carried out calculations with $\beta$=2 and 3.

We first point out that with $\beta=3$ and the SLy4 interaction we find $\rho^* > R_{WS}$ at the highest density,  so the criteria we explained in \ref{sec:method.pinningEnergy} are not met.
Therefore our method cannot produce a pinning energy value for that point.

Generally, the pinning energy has the same qualitative  behavior for both  interactions, with SkM* systematically predicting higher values.
At the lowest densities, the pinning energy is slightly negative and therefore nuclear pinning is favored.
On the other hand, the pinning energy grows considerably with $n_\infty$  up to about $n_\infty = 0.02$ fm$^{-3}$, implying that vortex lines are repelled at intermediate densities.
At the highest densities, the pinning energy either becomes roughly stable, as in the case of SkM*, or decreases, as for SLy4, where it even becomes negative again for $\beta=$ 2 and 3. 

At a given density the pinning energy decreases as a function of $\beta$. 
This can be understood, considering that the vortex radius (expressed in terms of its coherence length $\xi$) grows with $n_\infty$ and with $\beta$, as a larger value of $\beta$ corresponds to a lower pairing field $\Delta$.
We have previously seen that the vortex tends to incorporate the nucleus. 
This costs less energy if the vortex radius is larger, that is, for larger values of $\beta$, because the deformation needed is clearly less significant.
The nuclear pinning configuration, while still being not convenient, becomes less unfavorable and the pinning energy decreases considerably with $\beta$.

\begin{table}
   \centering
   \begin{ruledtabular}
   \begin{tabular}{cccc}
   $n_\infty $ [fm$^{-3}$]   &  \multicolumn{3}{c}{$E_p$ [MeV] (SLy4)}  \\
      &  {$\beta = 1$} &  {$\beta = 2$} & {$\beta = 3$}\\
      \hline
    $0.001$   &  $-0.72$  & $-0.48$  & $-0.27$    \\
    $0.002$   &  $-0.91$  & $-0.75$  & $-0.70$    \\
    $0.004$   &  $-0.89$  & $-0.97$  & $-0.93$    \\
    $0.008$   &  $2.73$   & $0.40$   & $-0.43$    \\
    $0.011$   &  $3.01$   & $0.63$   & $-0.26$    \\
    $0.017$   &  $10.00$  & $3.90$   & $1.06$    \\
    $0.026$   &  $11.78$  & $3.77$   & $-0.94$    \\
    $0.037$   &  $9.85$   & $-1.49$  & -          \\
   \end{tabular}
  \end{ruledtabular}
   \caption{Pinning energy and its uncertainty for eight different values of the neutron sea density.
            We show our results with the SLy4 interaction for the three different values of $\beta$.
            The highest density point with $\beta=3$ is absent because it does not satisfy our requirement $\rho^* > R_{WS}$ (see section \ref{sec:method.pinningEnergy}).}   
   \label{tab:pinningEnergiesSly}
   \end{table}

\begin{table}
   \centering
   \begin{ruledtabular}
   \begin{tabular}{cccc}
   $n_\infty $ [fm$^{-3}$]   &  \multicolumn{3}{c}{$E_p$ [MeV] (SkM*)}  \\
      &  {$\beta = 1$} &  {$\beta = 2$} & {$\beta = 3$}\\
      \hline
    $0.001$   &  $-0.19$  & $-0.30$  & $-0.27$    \\
    $0.002$   &  $-0.10$  & $-0.35$  & $-0.50$    \\
    $0.004$   &  $1.63$   & $0.18$   & $-0.23$    \\
    $0.008$   &  $7.47$   & $2.72$   & $1.19$    \\
    $0.011$   &  $8.06$   & $3.41$   & $1.68$    \\
    $0.017$   &  $11.12$  & $5.81$   & $3.59$    \\
    $0.026$   &  $19.07$  & $10.31$  & $6.47$    \\
    $0.037$   &  $18.69$  & $12.07$  & $6.43$    \\
   \end{tabular}
    \end{ruledtabular}
   \caption{Pinning energy and its uncertainty for eight different values of the neutron sea density.
            We show our results with the SkM* interaction for the three different values of $\beta$.}
   \label{tab:pinningEnergiesSkm}
   \end{table}

\begin{figure*}
    \centering
    \includegraphics[width=1\textwidth] {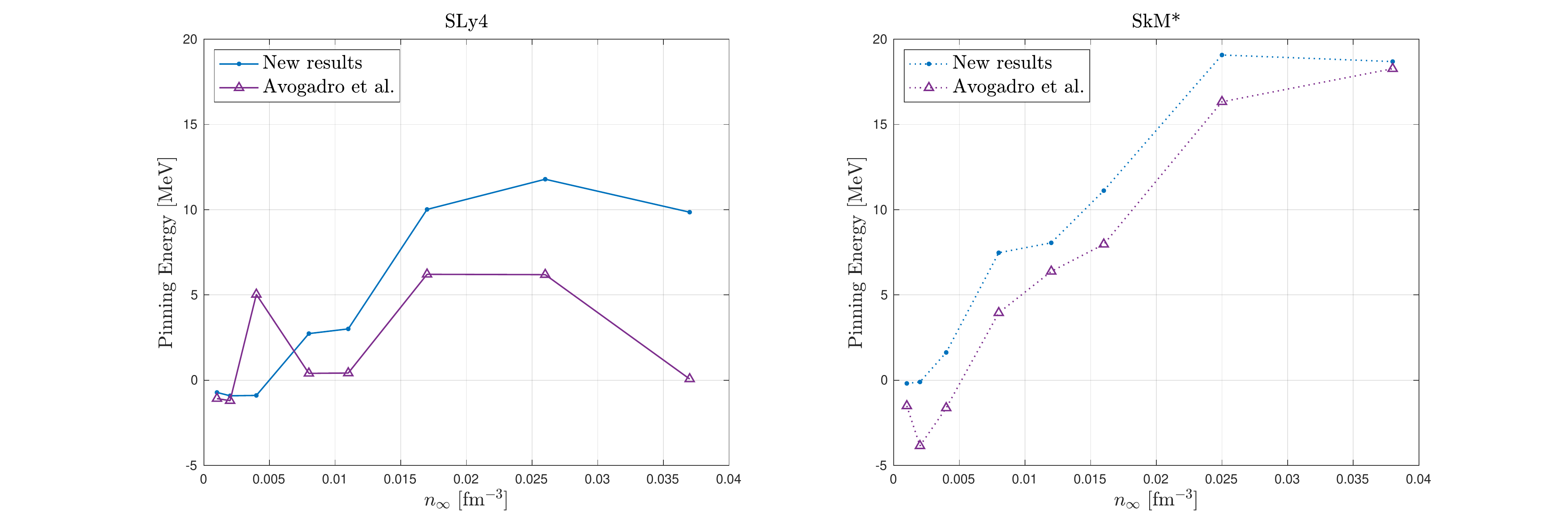}
    \caption{Comparison between our new results (blue dots) on the pinning energy and the results of Paper I \cite{Avogadro&Al} (purple triangles).
    As previously, we show the values as a function of the exterior neutron sea density $n_\infty$ for both Sly4 (left) and SkM* (right) interactions and for $\beta = 1$.}   
    \label{fig:confrontoAvogadro}
\end{figure*}
We carefully checked the dependence of our results on the radius of the Wigner-Seitz cell. 
We have found that generally, the computed pinning energies tend to stabilize for $R_{WS}$ larger than 35 fm.
For each set of parameters, we performed three calculations for $\rho_{WS}$= 38 fm, 40 fm, 42 fm, and the same height ($h_{WS} = 40$ fm). 
The resulting pinning energies differ by less than  $\sim10$ keV  at the lowest  density  we have considered, that is, $n_{\infty}$ = 0.001 fm$^{-3}$ and by less than 300 keV at $n_{\infty}$ = 0.017 fm$^{-3}$.
For a given density, we will report the  value averaged over the three boxes.
We have found that at the two largest computed densities, namely $n_{\infty}$ = 0.026 fm$^{-3}$ and $n_{\infty}$ = 0.037 fm$^{-3}$, the convergence pattern is more complicated, and we considered also larger values of $R_{WS}$, up to 48 fm. 
The HFB self-consistent process for the NP configurations can lead to two solutions having a different pairing and density spatial dependence, according to the box radius, and differing from each other by about 1.5 MeV.  
For these two densities, the boxes displaying the deepest minima were selected,  in keeping with the variational nature of our approach. 
The resulting uncertainty on the pinning energy is equal to  about 500 keV.



We conclude this section  comparing our results with those reported  in Paper I in Fig. \ref{fig:confrontoAvogadro}.
The pinning energies  computed with the SLy4 and the SkM* interaction are shown in the left and right panel respectively. 
Only the value $\beta = 1$ was considered in Paper I.
The results obtained for the SkM* interaction  are similar, aside from a sharp fall of the pinning energy in the second density zone.
On the other hand, for SLy4 the situation is rather different: the new results are more regular and grow monotonously  with $n_\infty$, while the previous ones present a distinct oscillatory behavior.
Quantitatively, the difference with the results of Paper I 
is substantial at the largest densities,
where the present pinning energies are larger by 5-10 MeV.
 
\begin{figure*}
    \centering
    \includegraphics[width=.8\textwidth] {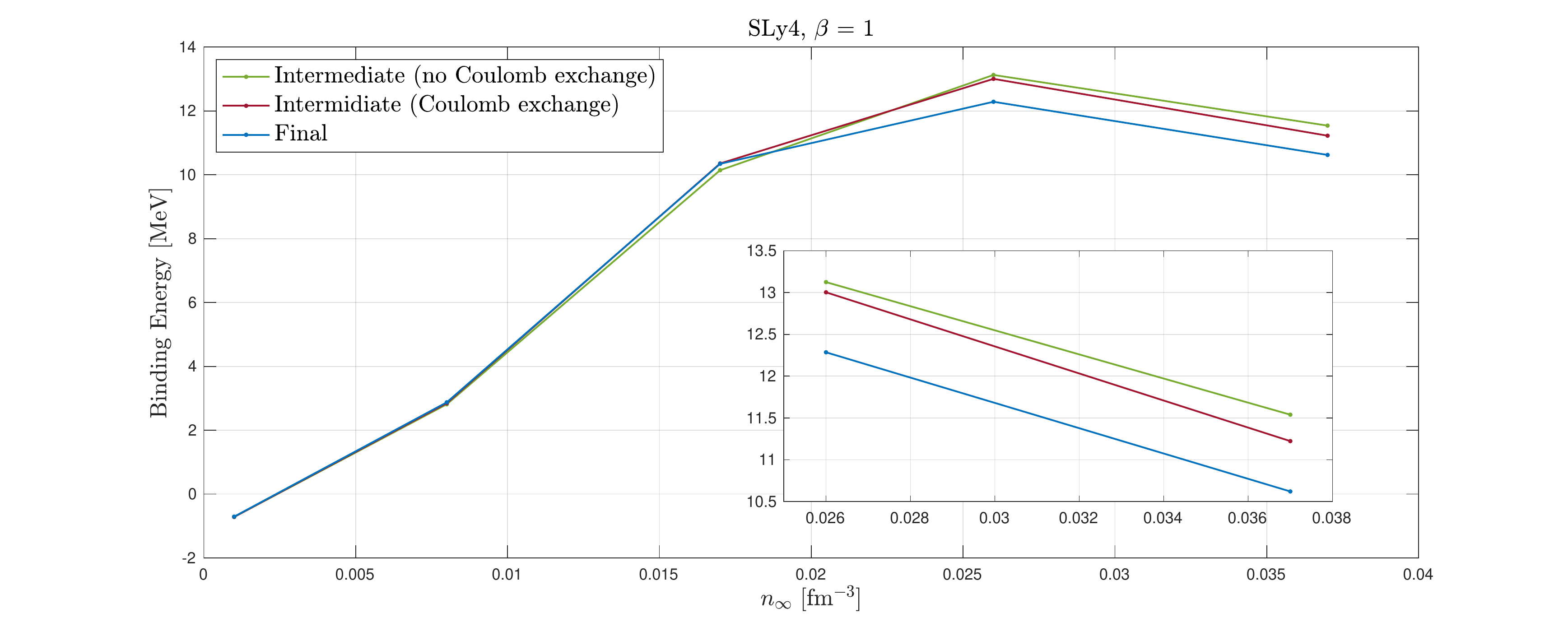}
    \caption{The pinning energy calculated with the SLy4 interaction for $\beta=1$ as a function of neutron density, already shown in Fig. \ref{fig:pinningEnergy}. 
    Our results (blue line) are compared with the one obtained neglecting both proton deformation and Coulomb exchange (green line) 
   or neglecting only proton deformation (red line).  
   The results obtained at the highest densities are shown in more detail in the inset.}   
    \label{fig:confrontoModifiche}
\end{figure*}

To study these differences in more detail, in Fig. \ref{fig:confrontoModifiche} we consider first the effect of  proton deformation and of Coulomb exchange, which were not taken into account in Paper I. 
  Proton deformation decreases the energy of 
the NP configuration; on the other hand, it does not affect the Nu configuration, in which we consider a spherical, closed shell nucleus. As a consequence (see Eq. \eqref{eq:bindingEnergyTheoric}) the pinning energy decreases, and therefore this effect cannot explain why the 
pinning energies are larger than those calculated in Paper I. 
In any case, one sees in  Fig. \ref{fig:confrontoModifiche} (see in particular the inset) that this effect is significant only for the largest densities, where it amounts to about 600-700 KeV. Neglecting deformation  but including Coulomb exchange, on the other hand,  decreases the pinning energy by at most about 100 keV.

We then conclude that the differences with Paper I must be  related to the improvements in the computational algorithms.
This point is further considered in Appendix B.




\subsection{Mesoscopic pinning forces}
\label{sec:meso}

The pinning energy contains information about the microscopic interaction between a vortex and a single nucleus.
Nonetheless, inner crust vortices are much longer than the lattice spacing and are expected to interact with many pinning sites \citep{PMP2016,AH_2020MNRAS}, giving rise to pinning at the mesoscopic scale (an intermediate scale in between the lattice spacing and the typical distance between two vortices in a pulsar).

\citet{PMP2016} found a simple prescription to estimate the mesoscopic pinning force per unit length $f_L$ acting on a vortex segment of length $L$, which is a better representative of the vortex-lattice interaction than the single-nucleus pinning energy, see the discussion in \citep{AH_2020MNRAS}.
They found an analytic approximation where the force per unit length $f_L = f_L (E_p,\,R_{WS},L)$ is a function of the pinning energy $E_p$ and the dimension of the WS cell $R_{WS}$.
This function depends also on the parameter $L$, the typical length over which a vortex filament in the inner crust could be approximated as straight. 
Finally, the estimate of $f_L (E_p,\,R_{WS},L)$ also depends on the geometrical properties of the lattice and on whether there is nuclear or interstitial pinning. However, the authors found that this distinction has a low impact on the pinning strength results, a result that is confirmed also by the dynamical simulations of an ensemble of vortices in complex pinning landscapes performed in \cite{AH_2020MNRAS,Link2022ApJ}.

\begin{figure*}
    \hspace*{-2cm}
    \includegraphics[width=.9\textwidth] {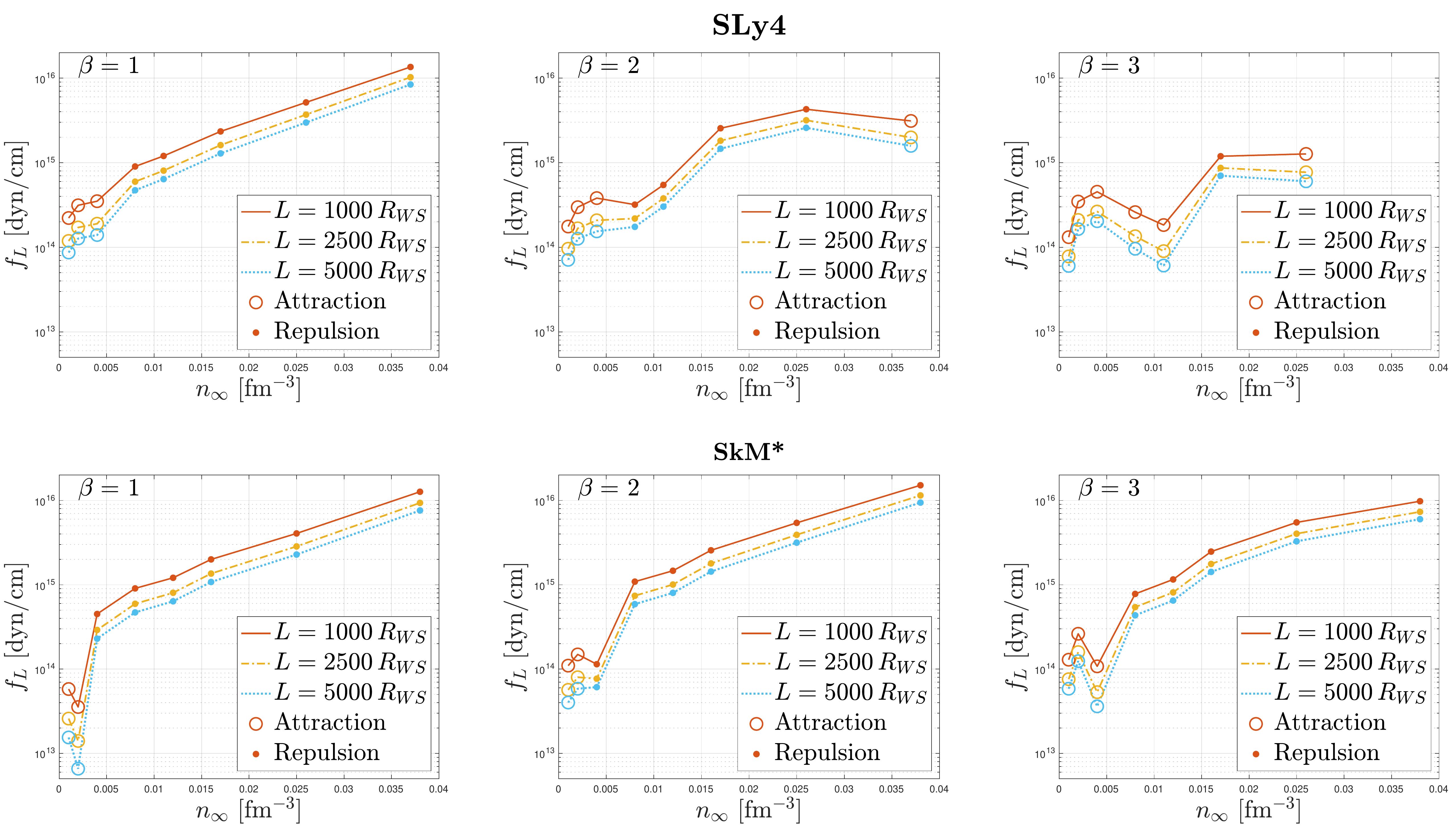}
    \caption{Absolute value of the pinning force per unit length as a function of the neutron sea density $n_\infty$, for both SLy4 (upper half) and SkM* (lower half) interactions. 
    Where it is attractive, we used a hollow circle, while where it is repulsive we used a dot.
    The values have been found using the prescription in \cite{PMP2016} for three different maximum-straight lengths $L = 1000$ (straight line), 2500 (line-dot), and 5000 (dotted line)  $R_{WS}$.
    We plotted the results for the three different values of $\beta$ used.
    As for the corresponding pinning energy, the highest density point with SLy4 and $\beta=3$ is absent because it does not satisfy our requirement $\rho^* > R_{WS}$ (see section \ref{sec:method.pinningEnergy}).}   
    \label{fig:pinningForce}
\end{figure*}

By following the procedure in \cite{PMP2016}, we can calculate new estimates for the typical pinning force for three different values of the parameter $L$ that defines the scale on which a vortex can be considered straight ($L = 1000, 2500, 5000 \; R_{WS}$, see  \cite{PMP2016}). 
Our results are shown in Fig. \ref{fig:pinningForce}.
We plot the absolute value of the force per unit length; where it is marked with dots, it is repulsive, otherwise, it is attractive where marked by circles. 
The mesoscopic pinning force values are of the same order of magnitude as the results of \cite{PMP2016}: the force per unit length ranges from $\sim 10^{13}$ dyn/cm up to $\sim 10^{16}$ dyn/cm.

While most of the remarks present in \cite{PMP2016} are valid for our results too, we briefly underline the following aspect.
The force decreases as the vortex length increases. 
Note that for an infinitely long and rigid vortex, the pinning force should vanish.
In fact, if the vortex were to move, the number of nuclei with which it interacts would not change\cite{jones,PMP2016}.


We can also compare our findings with the results of \cite{walzlowski2016}, which are obtained through a  different method.
In particular, from inset (b) of Fig.~3 of their work, we can see that they found a repulsive force of the order of $\sim 0.5$ MeV/fm when the vortex-nucleus distance is approximately 20 fm; after conversion to appropriate units, this is broadly consistent with our results.

\section{Conclusions} \label{sec:conslusions}

Microscopic pinning energies are a crucial ingredient in the dynamics of vortex-mediated pulsar glitches. The stronger the pinning of  a vortex line, the larger the amount of  angular momentum that can be stored in the inner crust in the form of a persistent (dissipationless) neutron current, which can then be potentially released in a glitch \cite{AMP_review_2023}.

Most of the past estimates of the pinning energies relied on a classical or semiclassical picture and had to use significant approximations to describe nuclei.
Working in the microscopic HFB framework solves these problems, as was done in Paper I \cite{Avogadro&Al}.  We have expanded and improved the latter work in four respects: we have 
i) allowed for the axial deformation of protons; 
ii) included the effect of the Coulomb exchange; 
iii) considered, although schematically, the effects of the screening of the pairing interaction;
and iv) improved the numerical treatment giving special attention to the convergence of our results. 
Based on these improvements, we found new and more reliable results on the pinning energy.

Our results show that nuclei attract vortices for the lower external neutron sea densities, while  the situation is the opposite at higher densities unless the pairing gap is strongly screened.
From our estimates of the pinning binding energy, we then extracted the typical force per unit length acting on a vortex, consistently with the procedure developed in \cite{PMP2016}. 
This force defines a theoretical upper limit on the depinning threshold \cite{AH_2020MNRAS} and, accordingly, an upper limit on the glitch amplitude in general relativity \cite{AMP_mnras_2018}. 
Therefore, in Sec. \ref{sec:meso} we have checked that our mesoscopic pinning forces are sufficiently large to be consistent with observations of giant glitches in the Vela pulsar. 

\acknowledgements{
    The Authors thank M. Antonelli for useful discussions and the careful reading of the manuscript including many useful suggestions. 
    F.~B. acknowledges the I+D+i project with Ref. PID2020-114687GB-I00, funded by MCIN/AEI/10.13039/501100011033.
}

\appendix
\section{Numerical details} \label{app:ND}
Within the HF approximation, one can obtain an explicit expression for the self-consistent potential of the Skyrme potential
\begin{equation}
    h(\mathbf{x}) = -\nabla \frac{\hbar^2}{2m^*_q(\mathbf{x})} \nabla + U_q(\mathbf{x}) + \delta_{q,p}V_C  
    \label{eq:hfHamiltonian}
\end{equation}
where $q$ can stand for $p$ (protons) or $n$ (neutrons).
Remembering that $n_q$ and $\tau_q$ are the density and the kinetic density of either protons or neutrons, and that $n=n_p+n_n$ and $\tau=\tau_p+\tau_n$, we write the terms in \eqref{eq:hfHamiltonian} following \cite{chabanat1}.
The effective mass $m^*_q$ is 
\begin{equation}
    \begin{aligned}
        \frac{\hbar^2}{2m^*_q(\mathbf{x})} = \frac{\hbar^2}{2m_q} + \frac{1}{8} \bigg[ t_1(2+x_1)+t_2(2+x_2)\bigg] n(\mathbf{x})   \\   
            - \frac{1}{8} \bigg[t_1(1+2x_1)+t_2(1+2x_2)\bigg] n_q(\mathbf{x})
    \end{aligned}
    \label{eq:effMass}
\end{equation}
the self-consistent potential $U_q$ reads
\begin{equation}
    \begin{split}
        U_q(\mathbf{x}) & = \frac{1}{2} t_0\bigg[ (2+x_0)n + (1+2x_0)n_q\bigg] \\
                        & + \frac{1}{24} t_3\bigg\{ (2+x_3)(2+\alpha)n^{\alpha+1} -\\
                        & (2x_3+1)\left[2n^\alpha n_q+\alpha n^{\alpha-1}(n_p^2+n_n^2)\right]\bigg\} \\
                        & + \frac{1}{8} \bigg[t_1(2+x_1)+t_2(2+x_2) \bigg] \tau + \\
                        & \frac{1}{8} \bigg[t_2(1+2x_2)-t_1(1+2x_1) \bigg] \tau_q \\
                        & + \frac{1}{16} \bigg[ t_2(2+x_2) - 3t_1(2+x_1)\bigg]\nabla^2n \\
                        & + \frac{1}{16} \bigg[ t_2(1+2x_2) + 3t_1(1+2x_1)\bigg]\nabla^2n_q \\
    \end{split}
    \label{eq:selfPotential}
\end{equation}
Lastly, the Coulomb potential, with the Slater approximation for the exchange part, reads
\begin{equation}
    V_C (\mathbf{x}) = e^2 \left( \int \frac{n_p(\mathbf{x}') \mathrm{d}_3x'}{|\mathbf{x}-\mathbf{x'}|}  -  \left(\frac{3}{\pi}\right)^{\frac{1}{3}}  n_p(\mathbf{x})^{\frac{1}{3}} \right)
    \label{eq:coulombPotential}
\end{equation}
In the code, we neglect the spin-orbit interaction, taking into account the spin simply with a degeneracy factor $g=2$. 


Each term of the potentials contributes to a term of the energy density of the system $\mathcal{H_{HF}}(\mathbf{x})$, which in turn is subdivided into different components
\begin{equation}
    \mathcal{H_{HF}} = \mathcal{K} + \mathcal{H}_0+ \mathcal{H}_3 + \mathcal{H}_{eff} + \mathcal{H}_{fin}+\mathcal{H}_C
    \label{eq:energyDensity}
\end{equation}
where each term reads
\begin{equation}
    \begin{split}
        \mathcal{K} & = \frac{\hbar^2}{2m} \tau \\
        \mathcal{H}_0 & = \frac{1}{4}t_0 \bigg[(2+x_0)n^2 - (2x_0+1)(n_p^2+n^2_n) \bigg] \\
        \mathcal{H}_3 & = \frac{1}{24} t_3 n^\alpha \bigg[ (2+x_3) n^2 - (2x_3 +1)(n^2_p + n^2_n) \bigg] \\
        \mathcal{H}_{eff} & = \frac{1}{8} \bigg[t_1(2+x_1)+t_2(2+x_2)\bigg]\tau n  \\
                          & + \frac{1}{8} \bigg[t_2(2x_2+1)-t_1(2x_1+1) \bigg] (\tau_pn_p +\tau_nn_n) \\
        \mathcal{H}_{fin} & = \frac{1}{32} \bigg[3t_1(2+x_1)- t2(2+x_2)\bigg] \left(\nabla n\right)^2  \\
                          & - \frac{1}{32} \biggl[3t_1(2x_1+1)+3t_2(2x_2+1)\biggr]  \left[\left(\nabla n_p\right)^2 + \left(\nabla n_n\right)^2\right]   \\
        \mathcal{H}_C & = e^2 \left( \frac{n_p}{2} \int\frac{n_p(\mathbf{x}') \mathrm{d}_3x'}{|\mathbf{x}-\mathbf{x'}|}  - \frac{3}{4} \left(\frac{3}{\pi}\right)^{\frac{1}{3}}  n_p(\mathbf{x})^{\frac{4}{3}} \right)                  
    \end{split}
    \label{eq:energyDensityComponents}
\end{equation}
We solve \eqref{eq:hfbCoordinate} in a cylindrical box with height $h_{box}$ and radius $\rho_{box}$.
We search for a solution expanded on a single-particle basis so that the amplitudes 
$u_{qm}(\rho,z,\varphi)$ and $v_{qm}(\rho,z,\varphi)$ for the quasi-particle level $q$ with projection of angular momentum along the $z$-axis $m$ are
\begin{equation}
    \begin{split}
        u_{qm}(\rho,z,\varphi) & = \displaystyle\sum_{nl} U_{qm}^{nl} f_{nm}(\rho)    g_l(z)e^{im\varphi} \\
        v_{qm}(\rho,z,\varphi) & = \displaystyle\sum_{nl} V_{qm}^{nl} f_{nm-\nu}(\rho)g_l(z)e^{i(m-\nu)\varphi} \\
    \end{split}
    \label{eq:uvExpantions}
\end{equation}

On the $\rho$ axis, functions $f_{nm}(\rho)$ are the solution of the Schr{\"o}dinger equation for free particles
\begin{equation}
    -\frac{\hbar^2}{2m_0} \left(  \frac{1}{\rho} \pdv{\rho}\left(\rho\pdv{\rho}\right) + \frac{m^2}{\rho^2} \right) f_{nm}(\rho) = e_{nm} f_{nm}(\rho)
    \label{eq:rhoAxis}
\end{equation}
where $m_0$ is the bare nucleon mass and the index $n$ is the number of nodes of function $f_{nm}(\rho)$ on the $\rho$ axis.

On the $z$ axis, functions $g_l(z)$ are normalized plane waves
\begin{equation}
    g_l(z) = \sqrt{\frac{2}{h_{box}}} \sin\left( k_l \left(z+\frac{h_{box}}{2}\right)\right), \; k_l = \frac{\pi}{h_{box}}, \frac{2\pi}{h_{box}}, \dots
    \label{eq:zAxis}
\end{equation}
so that we have
\begin{equation}
    \begin{split}
    -\frac{\hbar^2}{2m_0} \left( \pdv[2]{z} + \frac{1}{\rho} \pdv[2]{\varphi} + \frac{1}{\rho} \pdv{\rho}\left(\rho\pdv{\rho}\right)\right) & f_{nm}(\rho)g_l(z)e^{im\varphi} = \\
    \left(e_{nm}+\frac{\hbar^2k_l^2}{2m_0}\right)&f_{nm}(\rho)g_l(z)e^{im\varphi}       
    \end{split}
    \label{eq:freeSchrodinger}
\end{equation}

As for the boundary condition, each single-particle function vanishes at the edge of the box.

To solve \eqref{eq:hfbCoordinate}, we project it onto generic basis states $\ket{m_i,n_i,l_i} = \ket{\alpha_i}$.
Therefore our system of equations becomes, in matrix form
\begin{equation}
    \mqty (\matrixel{\alpha_2}{h - \lambda}{\alpha_1} & \matrixel{\alpha_2}{\Delta}{\alpha_1} \\
    \matrixel{\alpha_2}{\Delta^*}{\alpha_1} & -\matrixel{\alpha_2}{h - \lambda}{\alpha_1} )
    \label{eq:hfbProjection}
\end{equation}
Since $h$ depends only on the density, and the density does not depend on the azimuthal angle $\varphi$,
it holds 
\begin{equation}
    \matrixelement{m_2,n_2,l_2}{h}{m_1,n_1,l_1} = \delta_{m_1,m_2}\matrixelement{n_2,l_2}{h}{n_1,l_1}
    \label{eq:hfbHsymmetry}
\end{equation}
On the other hand, $\Delta = \Delta(\rho,z) e^{i\nu\varphi}$.
It follows 
\begin{equation}
    \matrixelement{m_2,n_2,l_2}{\Delta}{m_1,n_1,l_1} = \delta_{m_1,m_2+\nu}\matrixelement{n_2,l_2}{\Delta(\rho,z)}{n_1,l_1}
    \label{eq:hfbDsymmetry}
\end{equation}

We can now rewrite \eqref{eq:hfbCoordinate} explicitly. 
From \eqref{eq:hfHamiltonian} and \eqref{eq:pairingVortex}, we find
\begin{equation}
    \begin{cases}
        \displaystyle\sum_{n_2l_2} \left(h^{m}_{n_1l_1n_2l_2}-\lambda \right) U^{qm}_{n_2l_2} + \Delta^{m}_{n_1l_1n_2l_2}V^{qm}_{n_2l_2} & = E^{qm} U^{qm}_{n_1l_1} \\
        \displaystyle\sum_{n_2l_2} \Delta^{m}_{n_1l_1n_2l_2} U^{qm}_{n_2l_2} - \left(h^{m}_{n_1l_1n_2l_2}-\lambda \right)V^{qm}_{n_2l_2} & = E^{qm} V^{qm}_{n_1l_1}
    \end{cases}
    \label{eq:hfbTrue}
\end{equation}
where
\begin{widetext}
    
\begin{eqnarray}
        h^{m}_{n_1l_1n_2l_2} &=& 2\pi\int_0^{h_{box}} 2  \mathrm{d}z \; \int_0^{\rho_{box}} \rho\,\mathrm{d}\rho
        \Bigg\{ f_{n_2m}(\rho)g_{{l_2}}(z) \left( U(\rho,z) +\left(\frac{m_0}{m^*(\rho,z)}\right) \left( e_{n_1m} + \frac{\hbar^2 k_{l_1}^2}{2m_0} \right) -\lambda \right) f_{n_1m}(\rho) g_{{l_1}}(z) \nonumber \\
        &+& f_{n_2m}(\rho)g_{{l_2}}(z) \left( \pdv{\rho} \left(\frac{\hbar^2}{2m^*(\rho,z)} \right)  \cdot \pdv{f_{n_1m}(\rho)}{\rho}\right) g_{{l_1}}(z)
        + f_{n_2m}(\rho)g_{{l_2}}(z) \left( \pdv{z} \left(\frac{\hbar^2}{2m^*(\rho,z)} \right)   \cdot \pdv{g_{{l_1}}(z)}{z}\right) f_{n_1m}(\rho) \Bigg\} \nonumber\\
    \label{eq:hMatrixElement}
\end{eqnarray}
and
\begin{equation}
        \Delta^{m}_{n_1l_1n_2l_2} = 2\pi\int_0^{h_{box}} 2  \mathrm{d}z \; \int_0^{\rho_{box}} \rho\,\mathrm{d}\rho
        \left( f_{n_2m-\nu}(\rho)g_{{l_2}}(z) \Delta(\rho,z) f_{n_1m}(\rho) g_{{l_1}}(z)\right) 
    \label{eq:deltaMatrixElement}
\end{equation}
\end{widetext}

Since protons and neutrons feel different self-consistent potentials \eqref{eq:selfPotential}, they give rise to two systems \eqref{eq:hfbTrue}. 
From the solution of such systems, we then compute new densities, which we can use to write a new set of equations \eqref{eq:hfbTrue}. 
This iterative process stops once the relative energy difference between subsequent iterations is lower than an appropriate value.
Since protons are confined in the nucleus, the dimension of their box is smaller, fixed at 15 fm: so that it's big enough to contain all the protons but small enough to shorten the calculation times.
Finally, we do not consider proton pairing.
\section{Numerical Test} \label{app:40ca}

We test the accuracy of our axially deformed HFB code by 
applying it to the spherical nucleus $^{40}$Ca and comparing the results with the those obtained with the spherical code {\sc{hfbcs-qrpa}} \cite{hfrpa}.
For this test, we use the SLy4 interaction without the spin-orbit terms.

In Table \ref{tab:testNucleusEn} we show the total energy, divided among its contributions, as listed in \eqref{eq:energyDensityComponents};
the only exception being $E_{12}$, which is defined as $E_{12} = E_{fin} + E_{eff}$.
The relative difference between the {\sc{hfbcs-qrpa}} results and our program amount to 0.1-0.3\%.

In Table \ref{tab:testNucleusSPL} we list the single-particle energy levels of neutrons and protons.
We see that the present code reproduces the degeneracy of the 
Levels with the same values of the angular momentum $l$ within a few keVs, while deviations of the order of 100 keV are found in the original code.


\begin{table} [htb]
\caption{All contributions to the total energy ($E_{tot}$). Values are expressed in MeV. $\delta E$ is the relative energy difference (in percentage) between each value and the standard HF equivalent.
    We divided the energy in its main contributions, as in \eqref{eq:energyDensityComponents}, except for $E_{12}$, which is defined as $E_{12} = E_{fin} + E_{eff}$.
            The interaction used was SLy4 and the spin-orbit terms were neglected, as well as the Coulomb exchange potential.}
    \label{tab:testNucleusEn}
    
    \begin{ruledtabular}
    \begin{tabular}{lrrc}
    
    &	Ref. \cite{hfrpa} & {Present work} & $\delta E (\%)$ \\
\hline
    $K$ 		&$640.21$			&$638.93$		& 0.1	      	            \\
    $E_0$ 		&$-3716.80$			&$-3707.01$ 	& 0.3	  		\\
    $E_3$ 		&$2398.00$		    &$-2391.19$ 	& 0.3	  		\\
    $E_{12}$ 	&$279.59$    	    &$278.83$		& 0.3	      		          \\
    $E_C$ 		&$78.94$		    &$78.72$ 		& $0.3$	  		\\
    $E_{tot}$ 	&$-320.03$		    &$-319.33$ 		& $0.2$	  		\\
    \end{tabular}
    
    \end{ruledtabular}
   
    \end{table}
\begin{widetext}

\begin{table} [htb]
    \centering
    \begin{ruledtabular}
    \begin{tabular}{crcccc}
    \toprule
    & & \multicolumn{2}{c}{Neutrons} & \multicolumn{2}{c}{Protons} \\
    &  & Ref. \cite{hfrpa} &  Present work & Ref. \cite{hfrpa} & Present work \\
    \hline

    2s & $l_z = 0$ & $-16.95$  & $-16.889$ & $-9.48$  &  $-9.459$ \\
       & $l_z = 2$ & $-18.85$ & $-18.785$ & $-11.40$  & $-11.361$ \\
       & $l_z = 1$ & $-18.85$ & $-18.786$ & $-11.40$  & $-11.362$ \\
    1d & $l_z = 0$ & $-18.85$ &  $-18.789$ & $-11.40$ & $-11.371$ \\
       & $l_z =-1$ & $-18.85$ & $-18.786$ & $-11.40$ & $-11.362$ \\
       & $l_z =-2$ & $-18.85$ & $-18.785$ & $-11.40$ & $-11.361$ \\
       & $l_z = 1$ & $-33.21$ & $-33.184$ & $-25.29$ & $-25.282$ \\
    1p & $l_z = 0$ & $-33.21$ & $-33.182$ & $-25.29$ & $-25.277$ \\
       & $l_z =-1$ & $-33.21$ & $-33.184$ & $-25.29$ & $-25.282$ \\
    1s & $l_z = 0$ & $-47.82$ & $-47.799$ & $-39.36$ & $-39.356$ \\

    \end{tabular}
    \end{ruledtabular}
    \caption{Energies of each single particle level, both for protons and neutrons, expressed in MeV.}
    \label{tab:testNucleusSPL}
\end{table}
\end{widetext}

\section{$\rho^*$ criterion} \label{app:rhoStar}
We show here the values of the critical distance $\rho^* = R_N + \xi$ (see eq. \eqref{eq:rhoCrit}) for the two adopted Skyrme parametrizations  and for three values of the gap-reduction factor $\beta$.

{We observe that the value of $\rho^*$ is mostly determined by the pairing gap. 
As a consequence, $\rho^*$ has a minimum 
at intermediate densities, where the pairing gap reaches its maximum value.  

\begin{table} [htb]
    \centering
    \begin{ruledtabular}
    \begin{tabular}{ccccc}
    \toprule
    $n_\infty $ [fm$^{-3}$]  & $R_{WS}$ [fm] &  \multicolumn{3}{c}{$\rho^*$ [fm] (SLy4)}       \\
                             &               &  {$\beta = 1$} &  {$\beta = 2$} & {$\beta = 3$} \\
    \hline
    $0.001$                  &  $43.7$       & $11.8$         & $16.3$         &  $21.0$    \\
    $0.002$                  &  $41.5$       & $11.9$         & $16.0$         &  $20.0$    \\
    $0.004$                  &  $38.8$       & $11.4$         & $14.2$         &  $16.7$    \\
    $0.008$                  &  $33.7$       & $11.1$         & $13.1$         &  $14.9$    \\
    $0.011$                  &  $31.8$       & $11.2$         & $13.1$         &  $14.7$    \\
    $0.017$                  &  $28.9$       & $11.6$         & $13.6$         &  $15.3$    \\
    $0.026$                  &  $25.6$       & $12.5$         & $15.0$         &  $17.2$    \\
    $0.037$                  &  $21.4$       & $14.5$         & $18.5$         &  $21.7$    \\
       
    \end{tabular}
    \end{ruledtabular}
    \caption{Critical distance $\rho^*$ from our calculations with the SLy4 Skyrme parametrization.
    For $\beta = 3$, the value of $\rho^*$ is comparable to the dimension of the WS cell; therefore our method cannot estimate the pinning energy for this case.}
    \label{tab:testNucleusSPL}
\end{table}
\begin{table} [htb]
    \centering
    \begin{ruledtabular}
    \begin{tabular}{ccccc}
    \toprule
    $n_\infty $ [fm$^{-3}$]  & $R_{WS}$ [fm] &  \multicolumn{3}{c}{$\rho^*$ [fm] (SkM*)}       \\
                             &               &  {$\beta = 1$} &  {$\beta = 2$} & {$\beta = 3$} \\
    \hline
    $0.001$                  &  $43.7$       & $11.3$         & $15.9$         &  $20.6$    \\
    $0.002$                  &  $41.5$       & $11.7$         & $16.3$         &  $21.0$    \\
    $0.004$                  &  $38.8$       & $11.5$         & $13.9$         &  $19.7$    \\
    $0.008$                  &  $33.7$       & $10.7$         & $12.7$         &  $14.2$    \\
    $0.011$                  &  $31.8$       & $10.7$         & $12.4$         &  $13.9$    \\
    $0.017$                  &  $28.9$       & $10.7$         & $12.4$         &  $13.8$    \\
    $0.025$                  &  $25.6$       & $11.2$         & $12.9$         &  $14.3$    \\
    $0.038$                  &  $21.4$       & $12.3$         & $14.0$         &  $14.0$    \\
       
    \end{tabular}
    \end{ruledtabular}
    \caption{Critical distance $\rho^*$ from our calculations with the SkM* Skyrme parametrization.}
    \label{tab:testNucleusSPL}
\end{table}
\clearpage

\bibliography{bibliography}

\end{document}